\documentclass[runningheads]{llncs}

\usepackage{float}
\usepackage{graphicx} 
\usepackage{amsmath}
\usepackage{amssymb}
\usepackage{bbm}
\usepackage{mathtools}
\usepackage{makecell}
\usepackage{authblk}  
\usepackage[textsize=tiny]{todonotes}
\usepackage[hypertexnames=false,hyperfootnotes=false]{hyperref}
\usepackage{cleveref}
\usepackage{url}
\usepackage{ifthen}
\pdfoutput=1

\provideboolean{submissionversion}
\setboolean{submissionversion}{true}
\provideboolean{blindversion}

\ifthenelse{\boolean{submissionversion}}{
  \renewcommand{\todo}[2][1]{}
}{
}

\newtheorem{assumption}{Assumption}
\numberwithin{assumption}{section}
\numberwithin{theorem}{section}
\numberwithin{proposition}{section}
\numberwithin{corollary}{section}

\title{Quantifying the Value of Revert Protection}
\titlerunning{Revert Protection}

\ifthenelse{\boolean{blindversion}}{
  \author{Anonymous Authors}
  \authorrunning{Anonymous}
  \institute{}
}{
  \author{
    Brian Z. Zhu \inst{1} \and
    Xin Wan \inst{2} \and
    Ciamac C.\ Moallemi \inst{1,3} \and \\
    \vspace{-1em}
    Dan Robinson \inst{3} \and
    Brad Bachu \inst{2}
  }
  \institute{Columbia University \\
    \email{bzz2101@columbia.edu, ciamac@gsb.columbia.edu}
    \and
    Uniswap Labs \\
    \email{xin@uniswap.org, brad.bachu@uniswap.org}
    \and
    Paradigm \\
    \email{dan@paradigm.xyz}
  }
  \authorrunning{Zhu et al.}

 \date{Initial version: 16 October 2024}
}

\begin{document}

\maketitle

\begin{abstract}
    Revert protection is a feature provided by some blockchain platforms that prevents users from incurring fees for failed transactions. 
We study the economic implications and benefits of revert protection in the context of priority gas auctions and maximal extractable value. 
We develop a model in which searchers bid for a top-of-block arbitrage opportunity
under varying degrees of revert protection. 
This model applies to a broad range of settings, including bundle auctions on L1s and priority ordering sequencing rules on L2s.
We quantify, in closed form, how revert protection improves equilibrium auction revenue, market efficiency, and blockspace efficiency.



\end{abstract}

\section{Introduction and Background}

Revert protection is a feature for blockchains that block builders and sequencers can provide where they exclude
transactions that would otherwise fail, protecting users from paying fees for failed
transactions. However, failed transactions would pay fees, which often accrue to the block builder or sequencer. So is it in their interest to offer revert protection?

In this paper, we argue that revert protection is indeed in their interest and quantify the value of revert protection on blockchain platforms, particularly on the auctions that many of those
platforms use to capture and allocate maximal extractable value (MEV). We show that in various
relevant settings, revert protection is beneficial for auctioneer revenue, as well as for other
relevant outcomes like price discovery and blockspace efficiency.

The analysis is complicated by the wide range of rules --- across the complex Ethereum block builder market and on a variety of L2s --- for how transactions are selected for inclusion and ordered, how fees are collected, and who those fees go to, as well as how applications behave under those rules. We propose a model that can be parameterized to cover a wide variety of these settings, and use that model to solve for the equilibrium behavior of MEV searchers as a function of those parameters.

\subsection{Revert Protection}\label{sec:rp} On Ethereum and similar blockchains, users interact with smart contracts using atomic transactions. In the execution of a transaction, a smart contract may trigger a ``revert'' in the transaction, causing the entire transaction to fail. Normally, if the transaction is included, the user who sent the reverting transaction is still charged some transaction fee, even though the transaction has no other effect on the blockchain state. The fee charged is typically given by the product of \emph{gas price} of and the \emph{gas
  used} by the transaction before it either completes or reverts. The gas price can be
separated into the ``base fee''---a flat fee that must be paid by any
transaction in the block---and the ``priority fee''---an additional fee paid to the
builder that often affects inclusion and ordering (particularly when blocks are full). 

The gas used in the transaction is a function of how much of the transaction was executed---a transaction that reverts uses only some portion of the gas that it would have used had it succeeded. For example, on an automated market maker, typical reverting transactions may use only around 10–-20\% of the gas that a successful transaction would use. ``Revert protection'' is thereby the feature that block builders and sequencers may choose to implement in which failed transactions are excluded entirely from blocks. This feature improves user experience, as users only have to pay fees if their transaction succeeds. However, it also has significant effects on the behavior of the profit-seeking bots known as ``searchers,'' which we now describe.


\subsection{Block-Building Auctions}\label{sec:block-building}
Public blockchains have abundant opportunities for
MEV. As an example, one major type of MEV is the arbitraging of prices between decentralized and centralized
exchanges, also called ``CEX-DEX arbitrage,'' as discussed in \cite{daian2019flashboys20frontrunning} and \cite{milionis2022automated}. On these blockchains, independent actors known as ``searchers'' seek out MEV opportunities and compete to fill them.

Some block builders use different types of auctions to allocate these opportunities to searchers. For a given MEV opportunity, each transaction attempting to claim it can be thought of as a ``bid.'' Generally, only one transaction trying to claim a given opportunity will succed in extracting the value from it. Revert protection is therefore very relevant for these auctions, because it protects bidders on a given MEV opportunity from having to pay for failed bids if they do not win the opportunity.
Two examples of block building auctions include the ``bundle auctions'' run by builders such as Flashbots on Ethereum mainnet and the ``priority ordering'' rule implemented by sequencers on Ethereum L2s.

\paragraph{\textbf{Bundle Auctions.}} On Ethereum mainnet, blocks are typically built by profit-maximizing ``builders.'' Many of these builders run ``bundle auctions'' in which they allow any searcher to submit transactions, and use those searchers' bids as a factor when deciding which transactions to include and how to order them.\footnote{Block builders also play the role of bidders themselves in the ``MEV-Boost'' auction \cite{flashbots_auction}, in which they bid to have their block included by the current proposer. As ``bids'' are complete blocks in PBS auctions, revert protection is not as relevant here, and we will mostly set this case aside for the purposes of this paper.}

Bids can be expressed in two ways for this auction: (i) through the priority fee on the transaction or (ii) through a direct payment as part of the logic of the transaction by calling \verb|coinbase.transfer| \cite{flashbots_coinbase_transfer}.
The relevant difference between these methods of payment is that some portion of priority fees are paid \emph{even if the transaction reverts}, whereas if the transfer is made as part of the transaction, then it will be conditional on whether the transaction succeeds.

On Ethereum mainnet, the two components of gas price (base and priority fee) are paid in ETH and accrue to different users: base fees are burned, meaning they ultimately accrue to all ETH holders, rather than the builder; priority fees accrue to the block builder.\footnote{These bids, whether paid via priority fee or Coinbase transfer, technically go to the \emph{proposer}, not the builder who is assembling the block. However, since they reduce the amount that the builder has to pay to the proposer to win the MEV-Boost auction, we can think of these payments as a value transfer to the builder.} Many builders, such as Flashbots \cite{flashbots_protect}, provide revert protection for transactions submitted to them, even though it is not required by the Ethereum protocol and they might receive more in transaction fees by including it. Our results in this paper help explain this choice by showing that it likely increases their expected revenue in equilibrium.

\paragraph{\textbf{Priority Ordering.}} On Layer 2 blockchains today, blocks are typically built by a single sequencer, which often follows a deterministic algorithm for transaction inclusion and ordering. One of the most popular algorithms for this is ``priority ordering,'' in which transactions in a block are ordered in descending order of their gas price.

Priority ordering can be thought of as an auction in which transactions bid with the discretionary part of their gas price (i.e.\ their priority fee) to be included earlier in the chain. By default, this means that certain kinds of MEV (such as top-of-block CEX-DEX arbitrage) will generally accrue to the sequencer through priority fees. This means that when transactions fail, they still pay some of their priority fee to the sequencer. However, there is a mechanism called MEV taxes \cite{robinson_white_2024} that applications can use to capture all but a negligible portion of the value that would otherwise be paid through priority fees. Since MEV taxes are paid as part of the transaction and revert if the transaction fails, these fees will only be paid if the transaction succeeds.

\paragraph{\textbf{Considering All Cases.}} Even within these two settings of L1 block builders and L2 priority-ordered sequencers, we now need to consider at least seven cases that differ in who fees go to and how much is paid when the transaction reverts. Table 1 shows the differences between these cases. Our model is general enough to capture the distribution of transaction fees and revert behavior for all of these cases. We can compute in closed-form the expected base fee and remaining fee (i.e.\ priority fee, Coinbase transfer, or MEV tax) for all listed settings as well. For the remainder of the paper, we will use the terms ``builder'' and ``sequencer'' interchangeably.

\setlength{\tabcolsep}{0.25em} 

\begin{table}[h]
    \centering
    \begin{tabular}{|c|c|c|c|c|}
        \hline
        \textbf{Setting} & \makecell{\textbf{Base fee} \\ \textbf{goes to:}} & \makecell{\textbf{Rest of bid} \\ \textbf{goes to:}} & \makecell{\textbf{Is base fee} \\ \textbf{paid when} \\ \textbf{TX fails?}} & \makecell{\textbf{Is rest of bid} \\ \textbf{paid when} \\ \textbf{TX fails?}} \\ \hline
        \makecell{L1 block builder \\ with bids paid \\ via priority fees} & ETH holders & Builder & Partial & Partial \\ \hline
        \makecell{L1 block builder \\ with bids paid \\ via Coinbase transfer} & ETH holders & Builder & Partial & No \\ \hline
        \makecell{L2 sequencer with \\ priority ordering} & Sequencer & Sequencer & Partial & Partial \\ \hline
        \makecell{L2 sequencer with \\ priority ordering for \\ apps using MEV taxes} & Sequencer & Application & Partial & No \\ \hline
        \makecell{L1 block builder \\ with revert protection} & ETH holders & Builder & No & No \\ \hline
        \makecell{L2 sequencer with \\ revert protection} & Sequencer & Sequencer & No & No \\ \hline
        \makecell{L2 sequencer with \\ revert protection for \\ apps using MEV taxes} & Sequencer & Application & No & No \\ \hline
    \end{tabular} 
    \vspace{1em}
    \caption{Transaction Fee Distribution and Revert Behavior for Various Settings.}
    \label{tab:transaction_fee_distribution}
\end{table}

\subsection{Implementation Costs of Revert Protection} This paper is primarily concerned with the potential effects of revert protection, not with its implementation. We note that feasible revert protection is a difficult technical challenge: knowing whether a transaction will revert usually requires a builder to execute part of the transaction itself, which consumes computational resources. If the builder does not charge the sender anything for that reverted transaction, denial-of-service (DOS) attacks are possible by spamming the builder with an overwhelming number of reverting transactions.

There are some methods to mitigate this kind of attack. In some settings, the builder can use out-of-band spam prevention techniques (such as IP blocking) to make DOS attacks infeasible. For certain cases---including the auction-like use cases described above---it is also possible for the builder to determine statically whether a transaction will fail or not. 

\textcolor{black}{We abstract away from these challenges in our baseline model, assuming that the builder can implement revert protection with negligible cost. In our extensions, we study two schemes to deal with the implementation costs of revert protection where (i) costs are internalized by the sequencer and (ii) searchers pay a fixed amount for revert protection services.}


\subsection{Our Contributions}

The contributions of this paper are as follows:
\begin{itemize}
\item We introduce a novel and unified game-theoretic model that can be used to analyze revert
  protection in a variety of settings such as L1 block builders, L2 priority-ordered sequencers,
  MEV taxes, etc.
\item We characterize equilibria of our model in closed form, in terms of the model parameters: the
  value of the MEV opportunity, base fee, revert penalty parameters, and number of
  participating agents.
\item Using our model equilibrium, we can quantify the benefits of revert protection versus not
  offering revert protection:
  \begin{itemize}
  \item \textbf{Revert protection offers higher auction revenue.} The auctioneer extracts the full value of the MEV opportunity when the auction clears. However, in the
    absence of revert protection, agents randomize when they participate, and there is a non-zero
    probability that the auction does not clear and value is lost. This results in reduced
    sequencer revenue.
  \item \textbf{In the context of automated market makers, revert protection offers better market
    efficiency.} Here, the auction represents a CEX-DEX arbitrage opportunity. In the absense of
    revert protection, there is some chance that auctions do not clear, leaving arbitrage opportunities
    unexploited and hence prices less accurate.
  \item \textbf{Revert protection offers better blockspace efficiency.} With revert
    protection, only a single, winning transaction consumes block space. On the other hand,
    without revert protection, all submitted transactions consume block space.
  \end{itemize}
\item Our model allows for different revert penalty rates for the base fee and priority
  fees. While the penalty rate for priority fees influences bidder behavior, we show that it does not affect aggregate outcomes (e.g.\ revenue, number of submitted transactions).
  \item \textcolor{black}{When implementation costs are non-negligible but not too large, we show that charging a fixed fee to searchers for offering revert protection results in more auction revenue than builders internalizing the costs themselves.}
\end{itemize}

\subsection{Literature Review}

We contribute to the literature on on-chain MEV auctions, first explored in the seminal work of Daian et al.\ \cite{daian2019flashboys20frontrunning}. Studies have focused on aspects in the L1 block building environment, such as censorship resistance \cite{fox2023censorshipresistanceonchainauctions}, advantages for integrated searcher-builders \cite{paiMEVBoost2024}, private mempools \cite{CJWMEV}, and proposer-builder separation \cite{CJOPBS}. Other studies have focused on designing on-chain auctions and analyzing auction equilibria for L2 networks with various sequencing rules, including auctions where bidders are required to pay deposits \cite{schlegel2022onchain}, auctions under shared sequencing \cite{MSsharedsequencing}, taxes on MEV for applications \cite{robinson_white_2024}, and the TimeBoost mechanism \cite{MSbuyingtime} proposed for Arbitrum \cite{arbitrumwhitepaper} (a L2 network with a first-come first-serve sequencing rule). Relative to these studies, our paper offers a novel analysis on the effect of revert protection on on-chain auctions. Prior work has mentioned the congestion effects of reverting transactions in \cite{mevecosystem}, and the whitepaper for Unichain, an L2 network that plans to offer
revert protection, claims that it benefits swapping on decentralized exchanges
\cite{unichain}.

In the context of the economic literature on auctions, our model is a common-value all-pay auction with a reserve price and differential partial refunds for failed bids on the minimum bid and amount above the minimum bid. Common-value auction models are frequently used in the analysis of on-chain auctions \cite{CJOPBS,CJWMEV,MSsharedsequencing,paiMEVBoost2024} and in the traditional economics literature as well. Closest to our work is \cite{hillman1987dissipation}, which also considers a common-value all-pay auction with a reserve price, however does not allow for partial refunds on failed bids, and \cite{siegel2009}, whose model allows for a wide range of payoff and cost functions but has a more constrained strategy space for the bidders.



\section{Model}

\subsection{Auction Description} \label{auction-description}

We present a stylized model of a priority ordering auction for an MEV
opportunity. There are $N\geq 2$  agents, indexed by $i\in[N]$, bidding for a single MEV
opportunity in a block with common value $V > 0$, and the base gas fee for the block is $g >
0$.

Agent $i$ may choose to abstain or submit a bid (``priority gas fee'') $b_i\geq0$; we denote
the action of abstaining by $b_i = \varnothing$. The winner, denoted by $w$, extracts value $V$ from the
MEV opportunity and pays $g+b_{w}$. In the event of a tie, the winner is randomly selected among
the highest bids. Any losing agent $j$ does not receive any value and incurs a revert cost of
$r_1g+r_2b_j$, where $r_1,r_2\in[0,1]$ are the revert penalty rates on the base gas and priority
gas fees, respectively.\footnote{This model corresponds to a common value, all-pay auction with a
  minimum bid $g$ and differential refunds $(1-r_1) g$ and $(1-r_2) b_j$ for the base bid amount
  $g$ and additional bid amount $b_j$, respectively.}

\begin{table}[h]
    \centering
    \begin{tabular}{|c|c|c|}
        \hline
        \textbf{Setting} & \makecell{\textbf{Revert penalty} \\ \textbf{on base fee}} & \makecell{\textbf{Revert penalty} \\ \textbf{on rest of bid}} \\ \hline
      \makecell{L1 block builder with bids \\ paid via priority fees}
                         & \multicolumn{2}{c|}{$r_1 = r_2\in(0,1]$} \\ \hline
        \makecell{L1 block builder with bids \\ paid via Coinbase transfer} & $r_1\in(0,1]$ & $r_2=0$ \\ \hline
      \makecell{L2 sequencer with \\ priority ordering}
                         & \multicolumn{2}{c|}{$r_1 = r_2\in(0,1]$} \\ \hline
        \makecell{L2 sequencer with priority \\ ordering for apps using MEV taxes} & $r_1\in(0,1]$ & $r_2=0$ \\ \hline
      \makecell{L1 block builder \\ with revert protection}
                         & \multicolumn{2}{c|}{$r_1 = r_2 = 0$} \\ \hline
      \makecell{L2 sequencer with \\ revert protection}
                         & \multicolumn{2}{c|}{$r_1 = r_2 = 0$} \\ \hline
      \makecell{L2 sequencer with revert \\ protection for apps using MEV taxes}
                         & \multicolumn{2}{c|}{$r_1 = r_2 = 0$} \\ \hline
    \end{tabular}
    \vspace{1em}
    \caption{Revert Penalty Parameters for Settings in Table \ref{tab:transaction_fee_distribution}}
    \label{tab:transaction_fee_r1r2}
\end{table}

This model provides a unified setting that captures a variety of proposed and currently in-use
block building protocols among popular blockchains, including those discussed in
\Cref{sec:block-building} and listed in
\Cref{tab:transaction_fee_distribution}. \Cref{tab:transaction_fee_r1r2} illustrates how the
revert penalty parameters may be set in various settings. For example, with an L1 block builder
and bids paid via priority fee, we would expect $r_1 = r_2 > 0$, and these parameters might take a
value of 10--20\% for automated market maker swap transactions.

In the cases involving MEV taxes in \Cref{tab:transaction_fee_r1r2}, observe that $r_2 = 0$. This
is because with MEV taxes, applications can decide what fraction of the priority fees to capture
themselves versus giving to the sequencer. Priority fees that are captured by the application are
fully refunded on revert. As we will see later on (cf.\ \Cref{thm:revenue}), the
total priority fee revenue does not depend on the choice of $r_2$. Hence, applications are
incentivized to capture all but a negligible portion of the priority fee, and thus $r_2=0$. We make the following assumption to avoid trivialities:\footnote{This is because, if $V \leq g$,
  then the utility an agent receives when winning the action cannot be positive, even if the
  priority fee is zero, since the value does not exceed the base fee. Thus, all agents abstaining
  is a dominant strategy equilibrium.}
\begin{assumption}
  Assume that the value exceeds the base fee, i.e., $V > g$.
\end{assumption}

\subsection{Strategy Spaces and Payoffs}


\paragraph{Payoffs Under Pure Strategies.}
Under pure strategies, agent $i$ has strategy space $\mathcal{B}=\varnothing\cup[0,\infty)$ and chooses an action $b_i\in\mathcal{B}$. Given a strategy profile $b=(b_i)_{i\in[N]}$, the payoff of agent $i$, denoted $u_i$, is
\begin{align*}
    u_i(b_i|b_{-i}) = \begin{dcases}
      (V-g-b_i)\mathbb{P}(w=i|b) & \\
      \qquad -(r_1g+r_2b_i)\big(1-\mathbb{P}(w=i|b)\big) & \text{if $b_i\geq0$}, \\
        0& \text{if $b_i=\varnothing$,}
    \end{dcases}
\end{align*}
where $\mathbb{P}(w=i|b)$ denotes the probability of agent $i$ winning under the strategy profile
$b$, i.e.,
\begin{align*}
    \mathbb{P}(w=i|b) = \frac{\mathbbm{1}\{b_i=b_M\}}{1+\sum_{j\neq i}\mathbbm{1}\{b_j=b_i\}}.
\end{align*}
where $b_M\triangleq \max\{b_i \colon b_i \geq 0\}$, i.e., the highest participating bidder wins,
with ties broken at random. The first term in the payoff for participating bidders captures the
case when agent $i$ has the highest bid; their payoff is the net value gained, $(V-g-b_i)$, scaled
by the probability of winning. The second term captures the payment in the case when agent $i$
does not win the auction.

\paragraph{Payoffs Under Mixed Strategies.}  We now allow agents to probabilistically randomize
between the actions of abstaining from the auction and bidding a continuum of possible
values. Specifically, agent $i$ now chooses $\beta_i\equiv(p_i,F_i)$ where $p_i\in[0,1]$ is the
probability of abstaining ($b_i=\varnothing$), and $F_i$ is a continuous cdf supported on
$b_i \geq 0$ specifying the distribution that agent $i$ bids according to, conditional on agent
$i$ choosing to participate in the auction. Given a strategy profile $\beta=(\beta_i)_{i\in[N]}$,
the expected payoff $\bar u_i$ of agent $i$ conditional on realizing $b_i$ and assuming that
agents' actions are chosen independently, is
\begin{equation}\label{eq:u-1}
  u_i(b_i|\beta_{-i}) = \begin{dcases}
    (V-g-b_i)\mathbb{P}(w=i|b_i,\beta_{-i}) & \\
    \qquad-(r_1g+r_2b_i)\!\left(1-\mathbb{P}(w=i|b_i,\beta_{-i})\right) & \text{if $b_i \geq 0$,} \\
    0 & \text{if $b_i=\varnothing$,}
  \end{dcases}
\end{equation}
where $\mathbb{P}(w=i|b_i,\beta_{-i})$ is the probability that agent $i$ wins conditional on
realizing $b_i$ and the other agent's strategies. For the assumptions above, we have
\begin{equation}\label{eq:u-2}
  \mathbb{P}(w=i|b_i,\beta_{-i}) = \prod_{j\neq i}(p_j+(1-p_j)F_j(b_i)),
\end{equation}
noting that ties occur with probability zero under continuous distributions. The expected
utility $\bar u_i$ for agent $i$ over their random choice of action $b_i$ is then
\begin{equation}\label{eq:u-3}
  \bar u_i(\beta_i,\beta_{-i}) \triangleq \mathbb{E}\left[ u_i(b_i|\beta_{-i}) \right]
  = (1-p_i)\int_0^\infty u_i(b|\beta_{-i})\,dF_i(b).
\end{equation}





\section{Equilibrium}

\subsection{Pure Strategies}

A pure strategy profile $b^*$ as defined in the previous section is a Nash equilibrium in pure
strategies if no agent can unilaterally deviate to increase their payoff, i.e for all $i\in[N]$,
we have $u_i(b_i^*|b_{-i}^*) \geq u_i(b_i|b_{-i}^*)$ for any $b_i\in \mathcal{B}$.

\begin{theorem}\label{thm:pure-eq}
If $r_1=r_2=0$, then a pure strategy profile $b$ where bids are ordered such that $b_1 \leq b_2 \leq \dots \leq b_N$ (with abstentions represented by bids less than zero) is a Nash equilibrium if and only if $b_{N-1}=b_N=V-g$.
\end{theorem}

Intuitively, when there is no cost incurred upon losing the auction, the agents have no
disincentive associated with large bids, resulting in an equilibrium when at least two agents bid
the breakeven bid, i.e., the highest possible amount yielding a nonnegative utility, which is the
value of the arbitrage opportunity less the base gas fee.

\begin{theorem}\label{thm:no-eq}
If at least one of $r_1$ and $r_2$ is nonzero, then there does not exist a Nash equilibrium in pure strategies.
\end{theorem}

 Conversely, when agents incur any cost upon losing the auction, whether on the base or priority gas fee, there is no Nash equilibrium in pure strategies. The revert cost penalty results in a situation where at least one agent can benefit from unilaterally deviating given any pure strategy profile. Having characterized the pure-strategy Nash equilibrium and lack thereof, we now look at mixed-strategy equilibria.

\subsection{Mixed Strategies}

A mixed strategy profile $\beta^*$ as defined in the previous section is a Nash equilibrium if
$\bar u_i(\beta^*_i,\beta^*_{-i})\geq \bar u_i(\beta_i,\beta^*_{-i})$ for all agents $i$ and for any
other mixed strategy $\beta_i^*$. For tractability, we focus on solving for \textit{symmetric
  equilibria}, i.e., mixed-strategy Nash equilibria where all agents' strategies $\beta_i$ are
identical.

\begin{theorem}\label{thm:sym-mixed-eq}
If at least one of $r_1$ and $r_2$ is nonzero, then the unique symmetric mixed-strategy
equilibrium is given by
\begin{gather*}
    p_i^* = p^* \triangleq  \left(\frac{r_1g}{V-g+r_1g}\right)^{\frac{1}{N-1}},
   \\[0.25\baselineskip]
   F_i^* = F^*(b) \triangleq
     \frac{1}{1-p^*}\left(\left(\frac{r_1g+r_2b}{V-g-b+r_1g+r_2b}\right)^{\frac{1}{N-1}}-p^*\right),
\end{gather*}
for bids $b \in [0, V-g]$ and agents $i\in[N]$. The expected payoff of every agent in equilibrium is zero.
\end{theorem}

We remark that this equilibrium is unique among symmetric mixed-strategy Nash equilibria, but there may exist non-symmetric equilibria. A complete characterization of all possible equilibria of the auction is outside the scope of this paper, and we focus on the symmetric equilibrium for the rest of the paper.

In order for a mixed-strategy Nash equilibrium to arise, arbitrageurs must be indifferent over all possible actions supported by the mixed strategy. Thus, when the equilibrium assigns a positive probability mass to abstaining from the auction, an action which yields zero payoff, it follows that the equilibrium expected payoff of each arbitrageur is zero. It turns out that under the equilibrium in \Cref{thm:sym-mixed-eq}, equilibrium expected payoff for arbitrageurs is zero even in cases where they always choose to participate.

\begin{figure}[H]
    \centering
    \includegraphics[width=0.5\textwidth]{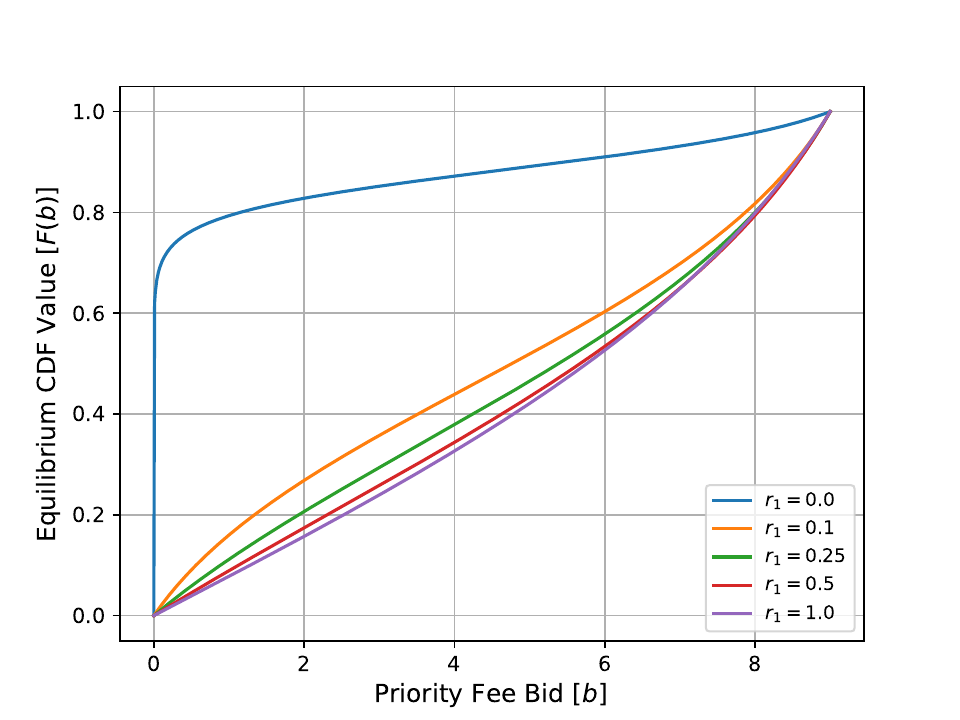} \hspace{-2.25em} \includegraphics[width=0.5\textwidth]{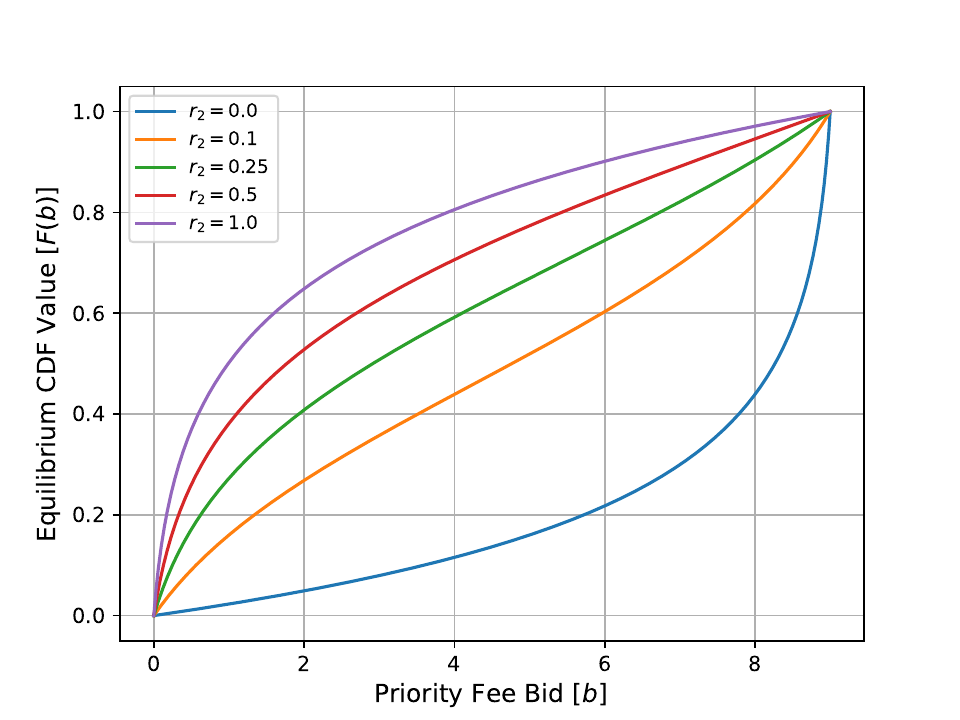} \\
    \caption{\centering Equilibrium CDF for Priority Fee Bids. We set $V=10$, $g=1$, $N=20$, $r_2=0.1$ for the left plot, and $r_1=0.1$ for the right plot.}
    \label{fig:cdf-r1-r2}
\end{figure}

\begin{figure}[H]
    \centering
    \includegraphics[width=0.5\textwidth]{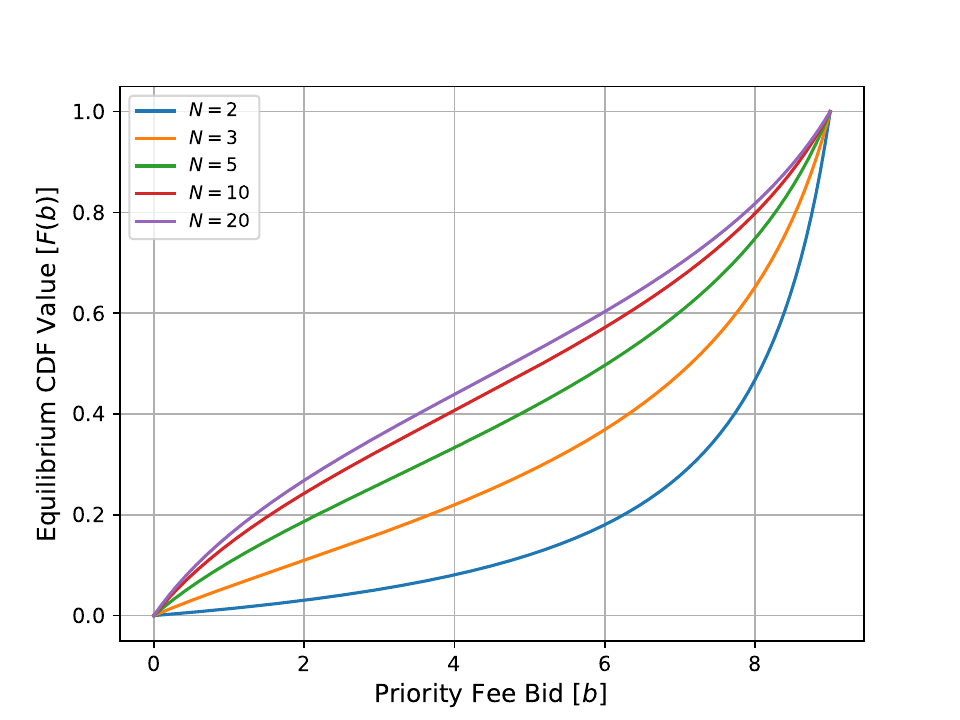} \hspace{-2.25em} \includegraphics[width=0.5\textwidth]{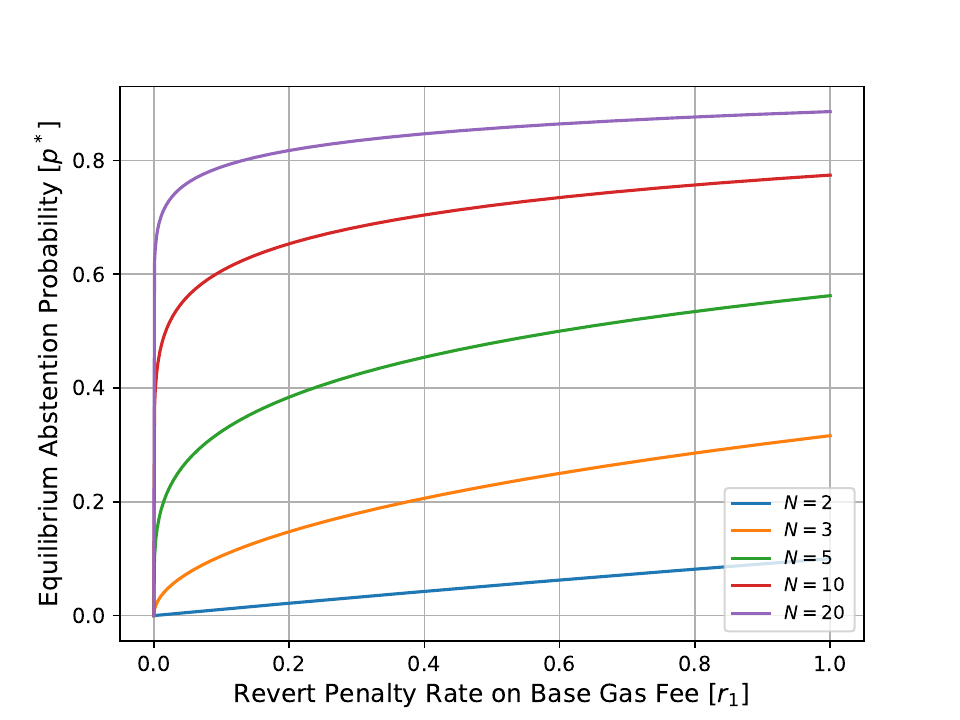} \\
    \caption{\centering Equilibrium CDF for Priority Fee Bids (left) and Abstention Probability (right). We set $V=10$, $g=1$, and $r_1=r_2=0.1$ for the left plot.}
    \label{fig:cdf-N-p}
\end{figure}

When $r_1>0$, the symmetric equilibrium strategy is characterized by a non-zero probability of
abstention and a non-degenerate continuous CDF specifying the distribution from which to draw the
priority gas fee bid. In the special case of $r_1=0$, all arbitrageurs participate with
probability one. Figures \ref{fig:cdf-r1-r2} and \ref{fig:cdf-N-p} plot the equilibrium CDF and
abstention probability for various parameters. \footnote{In \Cref{app:compstat}, we
discuss comparative statics for the the equilibrium CDF and
the abstention probability.}

We note that \Cref{thm:sym-mixed-eq} is consistent with \Cref{thm:pure-eq}. When
$r_1=r_2=0$, \Cref{thm:sym-mixed-eq} yields
$p^*=0$ and $F^*(b)=0$ for all $b\in[0,V-g)$. Compare this with the pure strategy equilibrium given in \Cref{thm:pure-eq},
which corresponds to a symmetric mixed-strategy Nash equilibrium given by $p^*=0$, $F^*(b)=0$ for
all $b\in[0,V-g)$, and $F^*(V-g)=1$. Indeed, one can show that $F^*(b)$ converges pointwise to the
function $\mathbbm{1}\{b=V-g\}$ as $(r_1,r_2)\to(0,0)$.

\section{Implications for Sequencer Design}\label{sec:implications}

In this section, we examine the comparative statistics of equilibrium quantities that are relevant to sequencer design.

\subsection{Auction Revenue and Market Efficiency}

\paragraph{\textbf{Auction Revenue.}} Given a realization of the mixed strategy, the auctioneer takes the base fee and priority fee bid of the winner (if one exists), and $r_1$ times the base fee plus $r_2$ times the priority fee bids of the remaining participating arbitrageurs. The revenue can be split into two terms: one capturing the revenue from the base fee, and the other capturing revenue from priority fee bids. The expected revenue over all possible realizations along with its decomposition into base and priority components is characterized by the following theorem.

\begin{theorem}\label{thm:revenue}
If at least one of $r_1$ and $r_2$ is nonzero, then under the symmetric mixed-strategy Nash
equilibrium:
\begin{itemize}
    \item The expected revenue from the auction is
    \begin{align*}
        \mathbb{E}[\mathsf{Revenue}] = \left(1-(p^*)^N\right)V = \left(1-\left(\frac{r_1g}{V-g+r_1g}\right)^{\frac{N}{N-1}}\right)V.
    \end{align*}
    \item Expected revenue decreases in $r_1$, does not depend on $N$ when $r_1=0$, decreases in $N$ when $r_1\neq0$, and does not depend on $r_2$.
    \item Expected revenue can be decomposed into components representing revenue from base gas fees and priority gas fees given by
    \begin{gather*}
        \mathbb{E}[\mathsf{BaseRevenue}] = \left(1-(p^*)^N\right)g+\left((1-p^*)N-\left(1-(p^*)^N\right)\right)r_1g, \\[0.25\baselineskip]
        \mathbb{E}[\mathsf{PriorityRevenue}] = \left(1-(p^*)^N\right)(V-g)-\left((1-p^*)N-\left(1-(p^*)^N\right)\right)r_1g.
    \end{gather*}
    \item If $r_1\neq0$, then as the number of arbitrageurs $N$ tends to infinity, the expected revenues converges to a finite limit, with
    \begin{gather*}
        \lim_{N\to\infty}\mathbb{E}[\mathsf{Revenue}] = \frac{V(V-g)}{V-g+r_1g}, \\[0.25\baselineskip]
        \lim_{N\to\infty}\mathbb{E}[\mathsf{BaseRevenue}] = \frac{(V-g)(g-r_1g)}{V-g+r_1g}-r_1g\log\left(1 + \frac{V-g}{r_1g}\right), \\[0.25\baselineskip]
        \lim_{N\to\infty}\mathbb{E}[\mathsf{PriorityRevenue}] = (V-g)+r_1g\log\left(1 + \frac{V-g}{r_1g}\right).
    \end{gather*}
\end{itemize}
\end{theorem}
The intuition behind the expression for total revenue is straightforward. Since the arbitrageurs
earn a combined expected payoff of zero in equilibrium and the winning arbitrageur extracts a
value of $V$, the total payments made to the sequencer should also equal $V$, as long as least one
arbitrageur participates. We then take a simple expectation over the events ``a value extraction
of $V$ occurs'' and ``no value extraction occurs'' which have probabilities $1-(p^*)^N$ and
$(p^*)^N$, respectively, for the result. This phenomena is known as ``rent dissipation''
\cite{hillman1987dissipation}: any value that is realized entirely goes to the sequencer.

Theorem \ref{thm:revenue} shows that expected revenue decreases in $r_1$, the revert penalty rate on the base gas fee. This implies that holding all other auction parameters equal, the revenue-maximizing choice of $r_1$ in expectation is zero, corresponding to a sequencer with full RP the base gas fee. Furthermore, expected revenue is constant in $N$ under full RP while it decreases in $N$ for nonzero $r_1$, highlighting additional losses when full RP is not implemented.

Interestingly, expected revenue does not depend on $r_2$, the revert penalty rate for priority gas fees. The intuition behind this is that value extraction only depends on participation, which is independent of $r_2$. Another interpretation is that $r_2$ does not matter for the ``marginal bidder'' who participates but bids a priority fee of zero. Consequently, sequencers and applications can adjust $r_2$ without affecting the expected total payments collected from participants.

This is relevant to considering MEV taxes imposed by applications, which not only change the
distribution of the non-base portion of the fee (causing a share of it to go to the application,
rather than the sequencer), but also affect $r_2$ (since MEV taxes are only paid when the
transaction succeeds, while a portion of priority fees are paid even on revert). Since $r_2$ does
not affect total revenue, this helps justify the assumption we made in Section
\ref{auction-description} that an application will parameterize its MEV taxes as high as possible
in order to maximize the revenue that it earns.

The expression for expected auction revenue coming from base gas fees is obtained by first conditioning on the number of participating arbitrageurs $k$. Given $k$ participants, the sequencer will earn $r_1g$ from $k-1$ of them and $g$ from the remaining one. Subtracting this from the total revenue thus gives the component coming from priority gas fee bids. Notably, these components also do not depend on $r_2$.




As previously mentioned, the probability of a value extraction occurring lowers as $N$ increases, stemming from the disincentive to participate brought on by a more competitive environment, but this probability converges to a finite limit as $N$ tends to infinity. This provides a minimum guarantee on expected revenue for any number of arbitrageurs.

\paragraph{\textbf{Market Efficiency.}} Revenue is deeply connected with market efficiency in automated market makers. Recall that the probability that value from the arbitrage opportunity is extracted is $1-(p^*)^N$. When this quantity is high, on-chain arbitrages are frequent, thereby improving market efficiency. On the other hand, when this quantity is low, on-chain arbitrages are less likely to occur, leaving arbitrage opportunities unexploited and generating zero value for the auction. It is straightforward to see that $r_1$ has the same directional impact on market efficiency (when measured by the probability that a value extraction occurs) as expected auction revenue described in Theorem \ref{thm:revenue}, so more revert protection (lower $r_1$) implies a higher market efficiency.

\subsection{Blockspace and Mempool Usage}

We proceed to analyze blockspace and memory pool usage. If an arbitrageur decides to participate, they submit a transaction to the mempool. When transactions are not fully revert-protected, they will still appear on-chain, which is prevented by full revert protection. The following theorem characterizes these quantities in equilibrium.

\begin{theorem}
If at least one of $r_1$ and $r_2$ is nonzero, then under the symmetric mixed-strategy Nash
equilibrium:
\begin{itemize}
    \item The expected number of submitted transactions is
    \begin{align*}
        \mathbb{E}[\mathsf{SubmittedTXs}]=(1-p^*)N = \left(1-\left(\frac{r_1g}{V-g+r_1g}\right)^{\frac{1}{N-1}}\right)N.
    \end{align*}
    \item Expected transactions submitted decreases in $r_1$ and does not depend on $r_2$.
    \item If $r_1\neq0$, then as the number of arbitrageurs $N$ tends to infinity, expected transactions submitted converge to a finite limit, with
    \begin{align*}
        \lim_{N\to\infty}\mathbb{E}[\mathsf{SubmittedTXs}]=\log\left(1 + \frac{V-g}{r_1g}\right).
    \end{align*}

\end{itemize}
\label{thm:blockspace}
\end{theorem}

Theorem \ref{thm:blockspace} implies that a higher revert penalty rate $r_1$ on the base gas fee has a dampening effect on arbitrageur participation, with the expected number of submitted transactions decreasing in $r_1$. Similarly to expected revenue, expected transactions submitted is constant in $r_2$ since submission only depends on the participation probability.

Under full revert protection, all arbitrageurs will participate, so as $N$ tends to infinity, so does the number of submitted transactions. When transactions are not fully protected, expected transactions submitted are bounded irrespective of the number of arbitrageurs. In terms of blockspace efficiency, full revert protection is more efficient, particularly as $V$ grows large, since only the winning arbitrageur's transaction will appear on the block as opposed to all participating arbitrageur's transactions when full RP is not in place. However, under full RP, there may be arbitrarily many transactions submitted to the mempool as $N$ grows large, compared to a bounded number of submitted transactions otherwise.


\subsection{Discussion: Full Revert Protection}

\begin{figure}[H]
    \centering
    \includegraphics[width=0.5\linewidth]{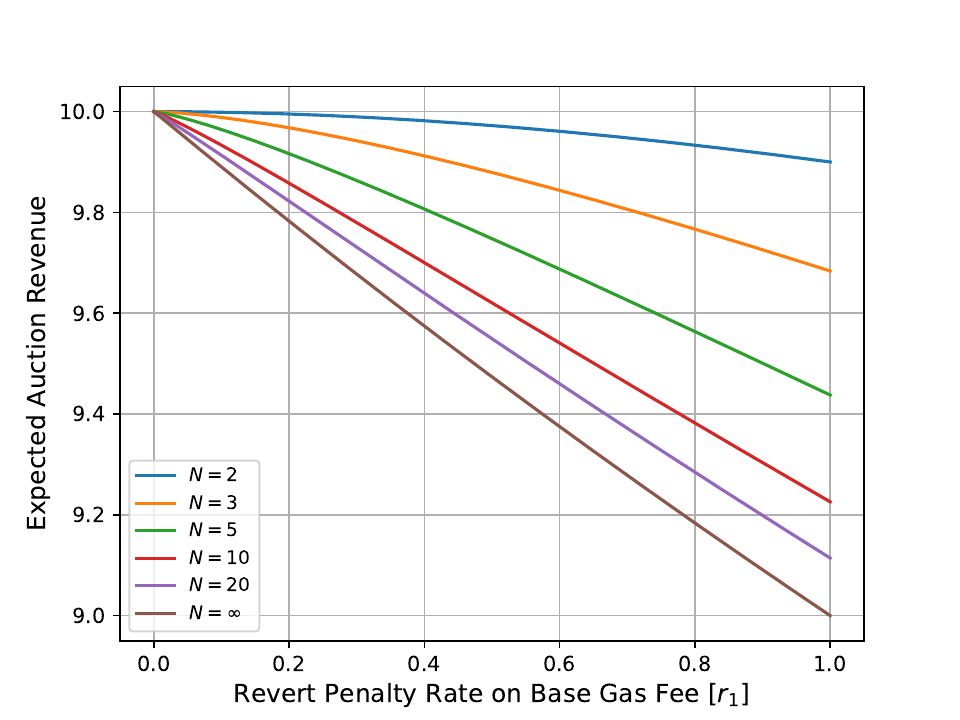} \hspace{-2.25em} \includegraphics[width=0.5\linewidth]{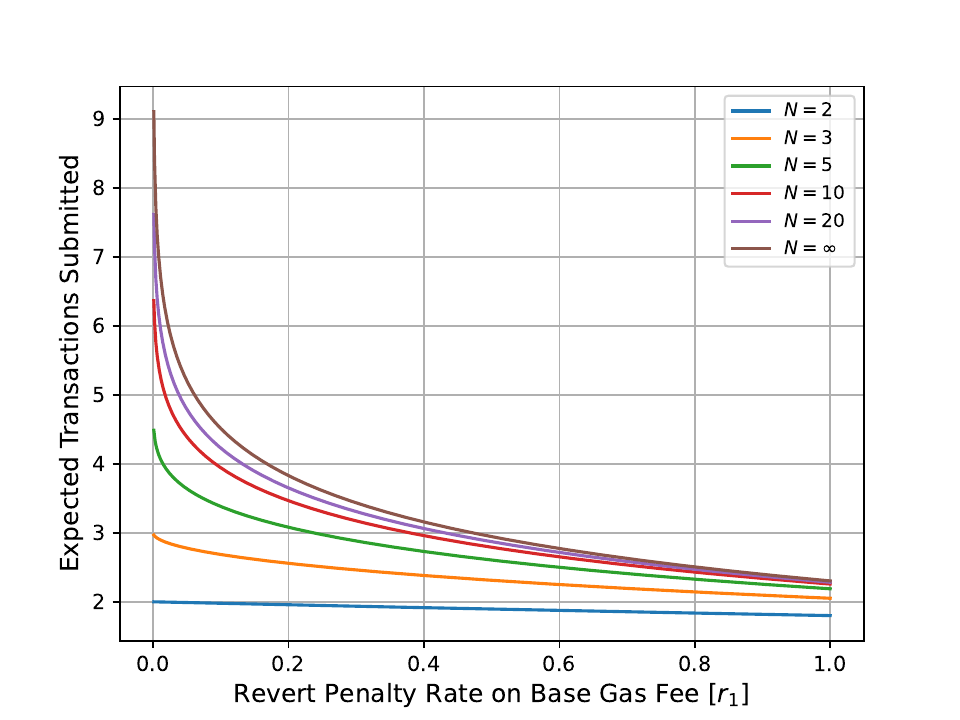} \\
    \caption{\centering Expected Auction Revenue (left) and Transactions Submitted (right). We set $V=10$ and $g=1$.}
    \label{fig:eq}
\end{figure}

We summarize the impact of full revert protection on arbitrage dynamics within a blockchain by analyzing and contrasting the expected number of participants and auction revenue in equilibrium under two conditions: one where $r_1\neq0$, and the other where $r_1=0$, i.e., full revert protection (for the base gas fee component). Figure \ref{fig:eq} shows these quantities for various values of $r_1$ and $N$.

With full revert protection, the sequencer is always able to capture the full value of an arbitrage opportunity, resulting in an expected auction revenue of $V$. In this scenario, since there is no deterrent to submitting bids, all arbitrageurs are incentivized to bid up to the breakeven point. As the number of arbitrageurs increases, this can present a spam risk to the sequencer. However, because reverts do not occur on-chain, at most one transaction appears on-chain when an arbitrage opportunity arises, or none otherwise. Thus, the primary challenge for the sequencer under a full revert protection is managing spam effectively.

When revert protection is not in place, the number of participants in an auction asymptotically approaches a finite number as the number of arbitrageurs grows indefinitely. Conversely, auction revenue is reduced by a factor of at most $ \left(1 - (p^*)^N\right) $ compared to the case of full revert protection. While full revert protection can enhance auction revenue and improve blockspace efficiency, it introduces the risk of increased spamming as the number of arbitrageurs grows. The central trade-off, therefore, lies between these benefits and the elevated risk of spam due to potentially unlimited bids as more arbitrageurs enter the system.

\section{Extensions}

So far, we assumed that revert protection can be implemented with negligible cost. This may not hold in practice, with block builders bearing non-negligible costs for implementing revert protection. We now take these costs into consideration and analyze two schemes in which costs are (i) internalized by the sequencer and (ii) passed on to searchers/arbitrageurs. In both scenarios, we now suppose that the cost of implementing RP per transaction submitted to the mempool is given by $c>0$. To avoid trivialities, we make the following assumption:
\begin{assumption}
Assume that the cost per transaction does not exceed the value less base fee, i.e., $c<V-g$.
\end{assumption}

\paragraph{\textbf{Scheme 1: Sequencer Internalizes Costs.}}

Here, the sequencer \textcolor{black}{incurs costs themselves}. While the equilibrium for the searchers does not change, the sequencer's expected revenue under internalization, denoted $\mathbb{E}[\textsf{Profit}]$, is 
\begin{align*}
    \mathbb{E}[\textsf{Profit}] &= \mathbb{E}[\textsf{Revenue}]-c\cdot\mathbb{E}[\mathsf{SubmittedTXs}] = (1-(p^*)^N)V-c(1-p^*)N.
\end{align*}
Since $p^*$ depends on $r_1$, the revert penalty rate on the base fee, we can optimize over $r_1$ to find the profit-optimal rate.

\begin{theorem}\label{thm:scheme1}
If at least one of $r_1$ and $r_2$ is nonzero, then under the symmetric mixed-strategy Nash equilibrium, the profit-optimal $r_1$ is
\begin{align*}
    r_1^* = \frac{c(V-g)}{(V-c)g}\cdot\mathbbm{1}\{c\leq g\}+\mathbbm{1}\{c>g\}
\end{align*}
\end{theorem}

Unlike the baseline model, where full revert protection maximized revenue, a nonzero revert penalty rate now maximizes profits when sequencers internalize the costs of implementing RP. This result highlights a tradeoff between offering full RP to searchers and optimizing profit while covering the costs of full RP.

\paragraph{\textbf{Scheme 2: Costs Passed Onto Searchers.}}

Here, searchers pay the builders an amount $c$ upfront for the cost of \textcolor{black}{simulating their transactions. The cost $c$ can also represent a ``cost of searching'' incurred by searching for MEV opportunities.} The searcher's payoff is 
\begin{equation*}\label{eq:u-1}
u_i(b_i|\beta_{-i}) = \begin{dcases}
(V-g-b_i)\mathbb{P}(w=i|b_i,\beta_{-i}) & \\
\qquad-(r_1g+r_2b_i)\!\left(1-\mathbb{P}(w=i|b_i,\beta_{-i})\right)-c & \text{if $b_i \geq 0$,} \\
0 & \text{if $b_i=\varnothing$.}
\end{dcases}
\end{equation*}
We resolve for the equilibrium and implications for sequencer design under the modified payoffs.

\begin{theorem}\label{thm:scheme2}
When searchers pay for the cost of revert protection
\begin{itemize}
    \item If $r_1=r_2=0$, then a pure strategy profile $b$ where bids are ordered such that $b_1 \leq b_2 \leq \dots \leq b_N$ (with abstentions represented by bids less than zero) is a Nash equilibrium if and only if $b_{N-1}=b_N=V-g-c$.
    \item If at least one of $r_1$ and $r_2$ is nonzero, then a pure-strategy Nash equilibrium does not exist. The unique symmetric mixed-strategy equilibrium is given by
    \begin{gather*}
    p_i^* = p^* \triangleq  \left(\frac{r_1g+c}{V-g+r_1g}\right)^{\frac{1}{N-1}},
       \\[0.25\baselineskip]
       F_i^* = F^*(b) \triangleq
         \frac{1}{1-p^*}\left(\left(\frac{r_1g+r_2b+c}{V-g-b+r_1g+r_2b}\right)^{\frac{1}{N-1}}-p^*\right),
    \end{gather*}
    for bids $b \in [0, V-g]$ and agents $i\in[N]$. The expected payoff of every agent in equilibrium is zero.
    \item Expected revenue decreases in $r_1$.
\end{itemize}
\end{theorem}

\paragraph{\textbf{Comparison of Schemes and Design Implications.}} Having analyzed the effects of revert protection costs on searchers and the sequencer in both schemes, we now compare the two and draw implications for sequencer design.

\begin{theorem}\label{thm:compare}
For $c$ sufficiently small, expected revenue in Scheme 2 is greater than the expected profit in Scheme 1 under the profit-optimal choice of $r_1$.
\end{theorem}

This result states that if the computational cost of implementing revert protection is sufficiently small, then having searchers pay the cost results in a greater expected profit to the sequencer than having the sequencer subsidize the costs itself. Besides yielding greater profits under small computational costs, the second scheme is also preferred in the sense that whereas the optimal $r_1$ needs to be adjusted based on auction parameters in the first scheme and is not necessarily zero, full revert protection is still optimal under the second scheme, with users paying a fixed cost rather than a penalty proportional to their bid, thus encouraging maximum participation in the PGA auction. 




\clearpage

\bibliographystyle{plainurl}
\bibliography{references}

\clearpage

\appendix
\section{Proofs}

\subsection{Proof of \Cref{thm:pure-eq}.}
Define $b^* \triangleq V -g$ to be the breakeven bid, and let $b$ be an ordered pure strategy
equilibrium.

Consider the following cases:
\begin{itemize}
\item $b_N > b^*$: In this case, the highest bidder \( b_N \) will incur a loss, as they are
  bidding more than the break-even bid $b^*$. This bidder would be better off abstaining,
  contradicting the fact that $b$ is an equilibrium.
\item \( b_N < b^* \): Here, the highest bidder \( b_N \) bids less than the break-even bid,
  allowing them to make a profit if they win. However, another bidder can deviate by bidding
  slightly above \( b_N \) but below \( b^* \), ensuring they win with certainty and still make a
  profit. Thus, the original set of bids cannot be in equilibrium, as there is an incentive to
  deviate.
\item \( b_N = b^* \) and \( b_{N-1} < b^* \): In this case, the top bidder is bidding exactly at
  the break-even level. However, the top bidder can increase their profit by reducing their bid to
  slightly above \( b_{N-1} \), as they would still win but at a lower bid cost. Hence, this setup
  also cannot be in equilibrium.
\end{itemize}

Conversely, suppose that $b_N = b_{N-1} = b^*$. Then, no matter the bids
of other players, all players have zero utility, and cannot increase their utility through any
deviation. Hence, this is an equilibrium.

\subsection{Proof of \Cref{thm:no-eq}.}

Suppose that $b$ is an pure strategy equilibrium, and assume the bids are ordered, so that
$b_1 \leq b_2 \leq \dots \leq b_N$, with the possibility of abstaining represented by bids less
than zero. Define the breakeven bid as $b^*\triangleq V - g$.

Consider the following cases:
\begin{itemize}
\item $b_N > b^*$: In this case, the highest bidder \( b_N \) will incur a loss, as they are
  bidding more than the break-even bid $b^*$. This bidder would be better off abstaining,
  contradicting the fact that $b$ is an equilibrium.
\item \( b_N < b^* \): Here, the highest bidder \( b_N \) bids less than the break-even bid,
  allowing them to make a profit if they win. However, another bidder can deviate by bidding
  slightly above \( b_N \) but below \( b^* \), ensuring they win with certainty and still make a
  profit. Thus, the original set of bids cannot be in equilibrium, as there is an incentive to
  deviate.
\item \( b_N = b^* \) and \( b_{N-1} < b^* \): In this case, the top bidder is bidding exactly at
  the break-even level. However, the top bidder can increase their profit by reducing their bid to
  slightly above \( b_{N-1} \), as they would still win but at a lower bid cost. Hence, this setup
  also cannot be in equilibrium.
\item $b_N = b_{N-1} = b^*$: If the top two bidders both bid the breakeven value \( b^* > 0 \), then
  they will make no profit when they win, and, because at least one of $r_1,r_2$ is positive, they
  will have strictly negative utility if they lose. Then, at
  least one player bidding $b^*$ will have strictly negative expected utility and would be better
  off abstaining. Therefore, this is not a sustainable equilibrium either.
\end{itemize}


\subsection{Proof of \Cref{thm:sym-mixed-eq}}

\todo{more careful proof, show it is unique}

We analyze a symmetric mixed strategy Nash equilibrium where arbitrageurs randomize their bids
over a range of possible values. Let \( p \) represent the probability that an arbitrageur
abstains from bidding, and for those who bid, let \( F(b) \) be the cumulative distribution
function (CDF) representing the probability that the bid is strictly less than \( b \).

Suppose an agent bids $b \geq 0$. From \eqref{eq:u-1}--\eqref{eq:u-3},
the expected probability of winning the auction is given by
\[
  \left( p + (1-p) F(b) \right)^{N-1},
\]
and thus the expected payoff is
\[
  \left( p + (1 - p) F(b) \right)^{N-1} (V - g - b)
  - \left( 1 - \left( p + (1 - p) F(b) \right)^{N-1} \right) (r_1g  + r_2 b).
\]
In a mixed strategy equilibrium, an arbitrageur must be indifferent between bidding and not
bidding. That is, the expected payoff from bidding any \( b \) must be zero, the same as the
expected payoff from abstaining. This leads to the following indifference condition
\[
\left( p + (1 - p) F(b) \right)^{N-1} (V - g - b) = \left( 1 - \left( p + (1 - p) F(b) \right)^{N-1} \right) (g \cdot r_1 + b \cdot r_2).
\]
Solving for $F(b)$, we have
\[
  F(b) =
  \frac{1}{1-p}\left(\left(\frac{r_1g+r_2b}{V-g-b+r_1g+r_2b}\right)^{\frac{1}{N-1}}-p\right).
\]
The boundary condition $F(0)=0$ yields
\[
F(0) =
\frac{1}{1-p}\left(\left(\frac{r_1g}{V-g-r_1g}\right)^{\frac{1}{N-1}}-p\right) = 0.
\]
Solving for $p$, we have that
\[
  p= \left(\frac{r_1g}{V-g+r_1g}\right)^{\frac{1}{N-1}}.
\]
Note that $F(V-g) = 1$, so that agents only bid in the range $b \in [0,V-g]$.

Now, assume that the agents all adopt the strategy $(p,F)$. We have established that any
individual agent is indifferent between abstaining, and any bid $b \in [0,V-g]$ --- all of these
actions result in zero expected payoff. Since bidding $b > V - g$ leads to a negative expected
payoff, such bids can be excluded. Therefore, the agents have no incentive to deviate, and we have
established a Nash equilibrium.

\subsection{Proof of \Cref{thm:revenue}}

\paragraph{Expected Revenue.} Each arbitrageur's expected utility can be decomposed into a ``value extracted'' and ``payment'' component:
\begin{align*}
    \bar u_i(\beta_i^*,\beta_{-i}^*) = \mathbb{E}[\mathsf{ValueExtracted}_i]-\mathbb{E}[\mathsf{Payment}_i].
\end{align*}
As each arbitrageur's expected utility is zero in equilibrium, summing all of them up yields
\begin{align*}
    0 &= \mathbb{E}\left[\sum_i\mathsf{ValueExtracted}_i\right]-\mathbb{E}\left[\sum_i\mathsf{Payment}_i\right] \\
    &=  \mathbb{E}[\mathsf{TotalValueExtracted}]-\mathbb{E}[\mathsf{Revenue}],
\end{align*}
so it follows that
\begin{align*}
     \mathbb{E}[\mathsf{Revenue}]=\mathbb{E}[\mathsf{TotalValueExtracted}].
\end{align*}
With probability $(p^*)^N$, no arbitrageurs participate, so no value is extracted. With probability $1-(p^*)^N$, at least one arbitrageur participates. In this case, only the winning arbitrageur will receive a value of $V$, with the others receiving zero value. Thus
\begin{align*}
     \mathbb{E}[\mathsf{Revenue}]=(p^*)^N\cdot0+(1-(p^*)^N)V=(1-(p^*)^N)V.
\end{align*}

\paragraph{Comparative Statics of Expected Revenue.} Note that
\begin{align*}
    \mathbb{E}[\mathsf{Revenue}]=1-(p^*)^N = 1-\left(\frac{r_1g}{V-g+r_1g}\right)^{\frac{N}{N-1}}.
\end{align*}
\begin{itemize}
    \item To show that $\mathbb{E}[\mathsf{Revenue}]$ decreases in $r_1$, note that
    \begin{align*}
        \frac{r_1g}{V-g+r_1g}
    \end{align*}
    increases in $r_1$.
    \item To show that $\mathbb{E}[\mathsf{Revenue}]$ decreases in $N$, note that
    \begin{align*}
        \frac{\partial}{\partial N}\left(\frac{r_1g}{V-g+r_1g}\right)^{\frac{N}{N-1}} = -\frac{1}{(N-1)^2}\left(\frac{r_1g}{V-g+r_1g}\right)^{\frac{N}{N-1}}\log\frac{r_1g}{V-g+r_1g}
    \end{align*}
    which is nonnegative since
    \begin{align*}
        \frac{r_1g}{V-g+r_1g}<1.
    \end{align*}
    \item It is straightforward to see that $\mathbb{E}[\mathsf{Revenue}]$ does not depend on $r_2$, and when $r_1=0$, we have $\mathbb{E}[\mathsf{Revenue}]=V$ which is constant in $r_2$.
\end{itemize}

\paragraph{Decomposition of Expected Revenue.} To derive the base gas fee revenue, let $N_P$ be a random variable for the number of agents that participate. Conditional on $N_P=k$ for $k\geq1$, note that the winner pays the full base fee $g$ while the $(k-1)$ losers pay the revert cost of $r_1g$. Then
\begin{align*}
    \mathbb{E}[\mathsf{BaseRevenue}] &= \sum_{k=1}^N\mathbb{P}(N_P=k)\cdot(g+(k-1)r_1g) \\[0.25\baselineskip] 
    &= r_1g\sum_{k=1}^Nk\cdot\mathbb{P}(N_P=k)+(1-r_1)g\sum_{k=1}^K\mathbb{P}(N_P=k) \\[0.25\baselineskip]
    &= N(1-p^*)r_1g+(1-(p^*)^N)(1-r_1)g.
\end{align*}
The revenue from priority fees is then given by
\begin{align*}
    \mathbb{E}[\mathsf{PriorityRevenue}] &= \mathbb{E}[\mathsf{Revenue}]-\mathbb{E}[\mathsf{BaseRevenue}] \\[0.25\baselineskip]
    &= (1-(p^*)^N)V-\left(N(1-p^*)r_1g+(1-(p^*)^N)(1-r_1)g\right).
\end{align*}

\paragraph{Limiting Values of Revenue Components.} Note that
\begin{align*}
    1-(p^*)^N = 1 -\left(\frac{r_1g}{V-g+r_1g}\right)^{\frac{N}{N-1}}. 
\end{align*}
As $N\to\infty$, note that $N/(N-1)\to1$, so
\begin{align*}
    \lim_{N\to\infty}1 -\left(\frac{r_1g}{V-g+r_1g}\right)^{\frac{N}{N-1}} = 1-\frac{r_1g}{V-g+r_1g}=\frac{V-g}{V-g+r_1g}.
\end{align*}
Similarly, note that
\begin{align*}
    (1-p^*)N = \left(1-\left(\frac{r_1g}{V-g+r_1g}\right)^{\frac{1}{N-1}}\right)N.
\end{align*}
As $N\to\infty$, the exponent $1/(N-1)$ tends to zero, so we have
\begin{align*}
    \lim_{N\to\infty}\left(1-\left(\frac{r_1g}{V-g+r_1g}\right)^{\frac{1}{N-1}}\right)N &= \lim_{N\to\infty}\left(-\frac{1}{N-1}\log\frac{r_1g}{V-g+r_1g}\right)N \\[0.25\baselineskip]
    &= \log\frac{V-g+r_1g}{r_1g}.
\end{align*}
Plugging these expressions into the formulas for expected revenue, expected base revenue, and expected priority revenue yields the result.

\subsection{Proof of \Cref{thm:blockspace}}

\paragraph{Expected Transactions Submitted.} For $i\in[N]$, let $Z_i$ be an indicator random variable for arbitrageur $i$ participating in the auction. Then
\begin{align*}
    \mathbb{E}[\mathsf{SubmittedTXs}] = \mathbb{E}\left[\sum_{i\in[N]}Z_i\right] = \sum_{i\in[N]}\mathbb{P}(Z_i=1) = N(1-p^*).
\end{align*}

\paragraph{Comparative Statics of Expected Transactions Submitted.} Note that
\begin{align*}
    \mathbb{E}[\mathsf{SubmittedTXs}] = (1-p^*)N = \left(1-\frac{r_1g}{V-g+r_1g}\right)^{\frac{1}{N-1}}N
\end{align*}
\begin{itemize}
    \item To show that $\mathbb{E}[\mathsf{SubmittedTXs}]$ decreases in $r_1$, note that
    \begin{align*}
        \frac{r_1g}{V-g+r_1g}
    \end{align*}
    increases in $r_1$.
    \item It is straightforward to see that $\mathbb{E}[\mathsf{SubmittedTXs}]$ does not depend on $r_2$.
\end{itemize}

\paragraph{Limiting Number of Expected Transactions Submitted.} The expression follows from the proof of \Cref{thm:revenue}.

\subsection{Proof of \Cref{thm:scheme1}}

The derivative of expected profit with respect to $r_1$ is 
\begin{align*}
    \frac{\partial\mathbb{E}[\textsf{Profit}]}{\partial r_1} = -N\cdot\frac{\partial p^*}{\partial r_1}\left((p^*)^{N-1}V-c\right)
\end{align*}
Since $N\cdot\frac{\partial p^*}{\partial r_1}>0$, dividing and equating to zero yields the FOC of
\begin{align*}
    (p^*)^{N-1}V = c \implies r_1^* = \min\left\{\frac{c(V-g)}{(V-c)g},1\right\} \implies p^* = \left(\frac{c}{V}\right)^{\frac{1}{N-1}}.
\end{align*}
Since $\frac{\partial\mathbb{E}[\textsf{Profit}]}{\partial r_1}$ at $r_1=0$ is positive and the FOC has at most one solution over $r_1\in[0,1]$, it follows that $r_1^*$ is indeed a maximizer of expected profits.

\subsection{Proof of \Cref{thm:scheme2}}

Similar to the proofs of \Cref{thm:pure-eq}, \Cref{thm:no-eq}, and \Cref{thm:sym-mixed-eq}.

\subsection{Proof of \Cref{thm:compare}}

Denote $\Pi_1$ and $\Pi_2$ as the expected profit in Scheme 1 under the optimal $r_1$ and expected revenue in Scheme 2 under $r_1=0$, respectively. It follows that 
\begin{gather*}
    \Pi_1 = \left(1-\left(\frac{c}{V}\right)^{\frac{N}{N-1}}\right)V-c\left(\left(\frac{c}{V}\right)^{\frac{1}{N-1}}\right)N, \ \Pi_2 = \left(1-\left(\frac{c}{V-g}\right)^{\frac{1}{N-1}}\right)V.
\end{gather*}

Taking derivatives with respect to $c$ yields
\begin{gather*}
    \frac{\partial\Pi_1}{\partial c} = -N\left(1-\left(\frac{c}{V}\right)^{\frac{1}{N-1}}\right), \ \frac{\partial^2\Pi_1}{\partial c^2} = \frac{1}{c(N-1)}\left(\frac{c}{V}\right)^{\frac{1}{N-1}}, \\[0.25\baselineskip]
    \frac{\partial\Pi_2}{\partial c} = -\frac{VN}{(V-g)(N-1)}\left(\frac{c}{V-g}\right)^{\frac{1}{N-1}}, \ \frac{\partial^2\Pi_2}{\partial c^2} = -\frac{1}{c(N-1)}\left(\frac{c}{V-g}\right)^{\frac{1}{N-1}}.
\end{gather*}
Note that $\Pi_1=\Pi_2$ when $c=0$. Since $\Pi_1$ is decreasing and convex in $c$ while $\Pi_2$ is decreasing and concave in $c$, there exists at most one other $c$ at which $\Pi_1$ and $\Pi_2$ intersect. Since $\frac{\partial\Pi_1}{\partial c}|_{c=0}=-N$ and $\frac{\partial\Pi_2}{\partial c}|_{c=0}=0$, we have $\Pi_2\geq\Pi_1$ initially.

\section{Comparative Statics of the Equilibrium}\label{app:compstat}

We examine how the equilibrium strategy changes with respect to the auction parameters $N$, $V$, $g$, $r_1$ and $r_2$. Specifically, we analyze the parameters' effect on the equilibrium abstention probability $p^*$ and the equilibrium priority gas fee bidding distribution $F^*$ via the expected bid $\mathbb{E}B_i^*$ where $B_i^*\sim F^*$ as a summary statistic.

\begin{theorem}
The equilibrium probability of abstention from the auction, $p^*$,
\begin{itemize}
    \item increases in $N$, $g$ and $r_1$;
    \item decreases in $V$.
\end{itemize}
The expected priority gas fee bid of an arbitrageur conditional on participation in the auction, $\mathbb{E}B_i^*$ where $B_i^*\sim F^*$,
\begin{itemize}
    \item increases in $V$;
    \item decreases in $N$, $r_1$, and $r_2$.
\end{itemize}
\label{thm:comp-stats}
\end{theorem}

These results are fairly intuitive. As the number of arbitrageurs increases, each arbitrageur has more competitors, reducing incentives to participate and bid higher in expectation due to the penalty associated with not winning. When the value of the arbitrage opportunity increases, the additional value incentivizes arbitrageurs to participate and submit higher expected bids. The effect of the base gas fee, $g$, on participation probability follows a similar intuition, but that on the expected priority fee bid is not characterizable in general without further specification of parameters.

As the revert penalty rate on the base gas fee increases, arbitrageurs are less inclined to participate and bid higher on average due tot he increase in revert costs. The revert penalty rate on the priority fee has a similar effect on the bidding distribution, but notably does not affect the participation probability at all. This is because a marginal arbitrageur who participates but bids zero priority fee is indifferent about the revert cost on priority fees.

\paragraph{Proof of Theorem \ref{thm:comp-stats}.}

Note that
\begin{gather*}
    \frac{\partial p^*}{\partial N} = -\frac{1}{(N-1)^2}\left(\frac{r_1g}{V-g+r_1g}\right)^{\frac{1}{N-1}}\log\frac{r_1g}{V-g+r_1g} \geq 0 \\[0.25\baselineskip]
    \frac{\partial p^*}{\partial g} = \frac{1}{N-1}\left(\frac{r_1g}{V-g+r_1g}\right)^{\frac{1}{N-1}}\frac{V}{g(V-g+r_1g)} \geq 0 \\[0.25\baselineskip]
    \frac{\partial p^*}{\partial r_1} = \frac{1}{N-1}\left(\frac{r_1g}{V-g+r_1g}\right)^{\frac{1}{N-1}}\frac{V-g}{r_1(V-g+r_1g)} \geq 0 \\[0.25\baselineskip]
    \frac{\partial p^*}{\partial V} = -\frac{1}{N-1}\left(\frac{r_1g}{V-g+r_1g}\right)^{\frac{1}{N-1}}\frac{1}{V-g+r_1g} \leq 0
\end{gather*}
where $p^*$ is the equilibrium abstention probability given by
\begin{align*}
    p^* = \left(\frac{r_1g}{V-g+r_1g}\right)^{\frac{1}{N-1}}.
\end{align*}
Note that
\begin{gather*}
    \frac{\partial F^*}{\partial N} = -\frac{\left(\frac{r_1g+r_2b}{V-g-b+r_1g+r_2b}\right)^{\frac{1}{N-1}}}{(1-p)(N-1)^2}\log\left(\frac{r_1g+r_2b}{V-g-b+r_1g+r_2b}\right) \geq 0 \\[0.25\baselineskip]
    \frac{\partial F^*}{\partial V} = -\frac{\left(\frac{r_1g+r_2b}{V-g-b+r_1g+r_2b}\right)^{\frac{1}{N-1}}}{(1-p)(N-1)(V-g-b+r_1g+r_2b)}\leq0 \\[0.25\baselineskip]
    \frac{\partial F^*}{\partial r_1} = \frac{g(V-g-b)\left(\frac{r_1g+r_2b}{V-g-b+r_1g+r_2b}\right)^{\frac{1}{N-1}}}{(1-p)(N-1)(r_1g+r_2b)(V-g-b+r_1g+r_2b)}\geq0 \\[0.25\baselineskip]
    \frac{\partial F^*}{\partial r_2} = \frac{b(V-g-b)\left(\frac{r_1g+r_2b}{V-g-b+r_1g+r_2b}\right)^{\frac{1}{N-1}}}{(1-p)(N-1)(r_1g+r_2b)(V-g-b+r_1g+r_2b)}\geq0
\end{gather*}
where $F^*$ is the equilibrium CDF given by
\begin{align*}
    \frac{1}{1-p^*}\left(\left(\frac{r_1g+r_2b}{V-g-b+r_1g+r_2b}\right)^{\frac{1}{N-1}}-p^*\right).
\end{align*}
If $\partial F^*/\partial\theta\geq0$ where $\theta$ is some parameter of interest, then the CDF increases pointwise for all $b\in[0,V-g]$. Then for any $\theta_1,\theta_2$ such that $\theta_1<\theta_2$, letting $B^*_{\theta_1}\sim F^*_{\theta_1}$ and $B^*_{\theta_2}\sim F^*_{\theta_2}$, it follows that $B^*_{\theta_1}$ stochastically dominates $B^*_{\theta_2}$ in the first order, so $\mathbb{E}B^*_{\theta_1}>\mathbb{E}B^*_{\theta_2}$.

If $\partial F^*/\partial\theta\leq0$ where $\theta$ is some parameter of interest, then the CDF decreases pointwise for all $b\in[0,V-g]$. Then for any $\theta_1,\theta_2$ such that $\theta_1<\theta_2$, letting $B^*_{\theta_1}\sim F^*_{\theta_1}$ and $B^*_{\theta_2}\sim F^*_{\theta_2}$, it follows that $B^*_{\theta_2}$ stochastically dominates $B^*_{\theta_1}$ in the first order, so $\mathbb{E}B^*_{\theta_1}<\mathbb{E}B^*_{\theta_2}$.










\end{document}